%% file: main.tex
\documentclass[sigconf]{acmart}

\AtBeginDocument{%
  \providecommand\BibTeX{{%
    \normalfont B\kern-0.5em{\scshape i\kern-0.25em b}\kern-0.8em\TeX}}}

\setcopyright{acmcopyright}
\copyrightyear{2024}
\acmYear{2024}
\setcopyright{acmlicensed}\acmConference[WWW '24 Companion]{Companion Proceedings of the ACM Web Conference 2024}{May 13--17, 2024}{Singapore, Singapore}
\acmBooktitle{Companion Proceedings of the ACM Web Conference 2024 (WWW '24 Companion), May 13--17, 2024, Singapore, Singapore}
\acmDOI{10.1145/3589335.3648319}
\acmISBN{979-8-4007-0172-6/24/05}

\usepackage{multirow}
\usepackage{marvosym}
\usepackage{diagbox}
\usepackage{graphicx}
\usepackage{balance}
\usepackage{hyperref}
\usepackage{hyperxmp}
\usepackage[inkscapelatex=false]{svg}
\begin{document}



\title{IU4Rec: Interest Unit-Based Product Organization and Recommendation for E-Commerce Platform}



\author{Wenhao Wu, Xiaojie Li, Lin Wang, Jialiang Zhou, Di Wu, Qinye Xie, Qingheng Zhang, Yin Zhang, Shuguang Han, Fei Huang, Junfeng Chen}
\affiliation{\{wenhaowu, ximo.lxj, wanglin.wang, zhoujialiang.zjl, wd442911, qinye.xqy, qingheng.zqh, jianyang.zy, shuguang.sh, huangfei.hf, jufeng.cjf\}@alibaba-inc.com
\country{}
}
\affiliation{%
  \institution{Alibaba Group}
  \city{Hangzhou}
  \country{China}
}


\renewcommand{\shortauthors}{Wenhao Wu et al.}
\begin{abstract}
Most recommendation systems typically follow a product-based paradigm utilizing user-product interactions to identify the most engaging items for users.
However, this product-based paradigm has notable drawbacks for Xianyu~\footnote{Xianyu is China's largest online C2C e-commerce platform where a large portion of the product are post by individual sellers}. Most of the product on Xianyu posted from individual sellers often have limited stock available for distribution, and once the product is sold, it's no longer available for distribution. This result in most items distributed product on Xianyu having relatively few interactions, affecting the effectiveness of traditional recommendation depending on accumulating user-item interactions.
To address these issues, we introduce \textbf{IU4Rec}, an \textbf{I}nterest \textbf{U}nit-based two-stage \textbf{Rec}ommendation system framework. We first group products into clusters based on attributes such as category, image, and semantics. These IUs are then integrated into the Recommendation system, delivering both product and technological innovations. 
IU4Rec begins by grouping products into clusters based on attributes such as category, image, and semantics, forming Interest Units (IUs). Then we redesign the recommendation process into two stages. In the first stage, the focus is on recommend these Interest Units, capturing broad-level interests. In the second stage, it guides users to find the best option among similar products within the selected Interest Unit. User-IU interactions are incorporated into our ranking models, offering the advantage of more persistent IU behaviors compared to item-specific interactions. This interest unit based recommendation can be beneficial from mitigating the side effect of limited-stock problem, since most interaction can be gathered on interest units which can persist and accumulate over time. 
Experimental results on the production dataset and online A/B testing demonstrate the effectiveness and superiority of our proposed IU-centric recommendation approach. 
This study not only advances recommendation technologies but also emphasizes the potential for co-evolution between product innovations and the technologies involved in item supply and distribution.

\end{abstract}





\maketitle

\input{tex/1_introduction.tex}
\input{tex/2_related_work.tex}
\input{tex/3_preliminaries.tex}

\input{tex/4_methodology.tex}
\input{tex/5_experiments.tex}

\input{tex/6_conclusions.tex}


\bibliographystyle{ACM-Reference-Format}
\balance
\bibliography{main}



\end{document}

%% file: tex/1_introduction.tex
\section{Introduction}



\begin{figure*}[tbp]
    \includegraphics[width=17cm]{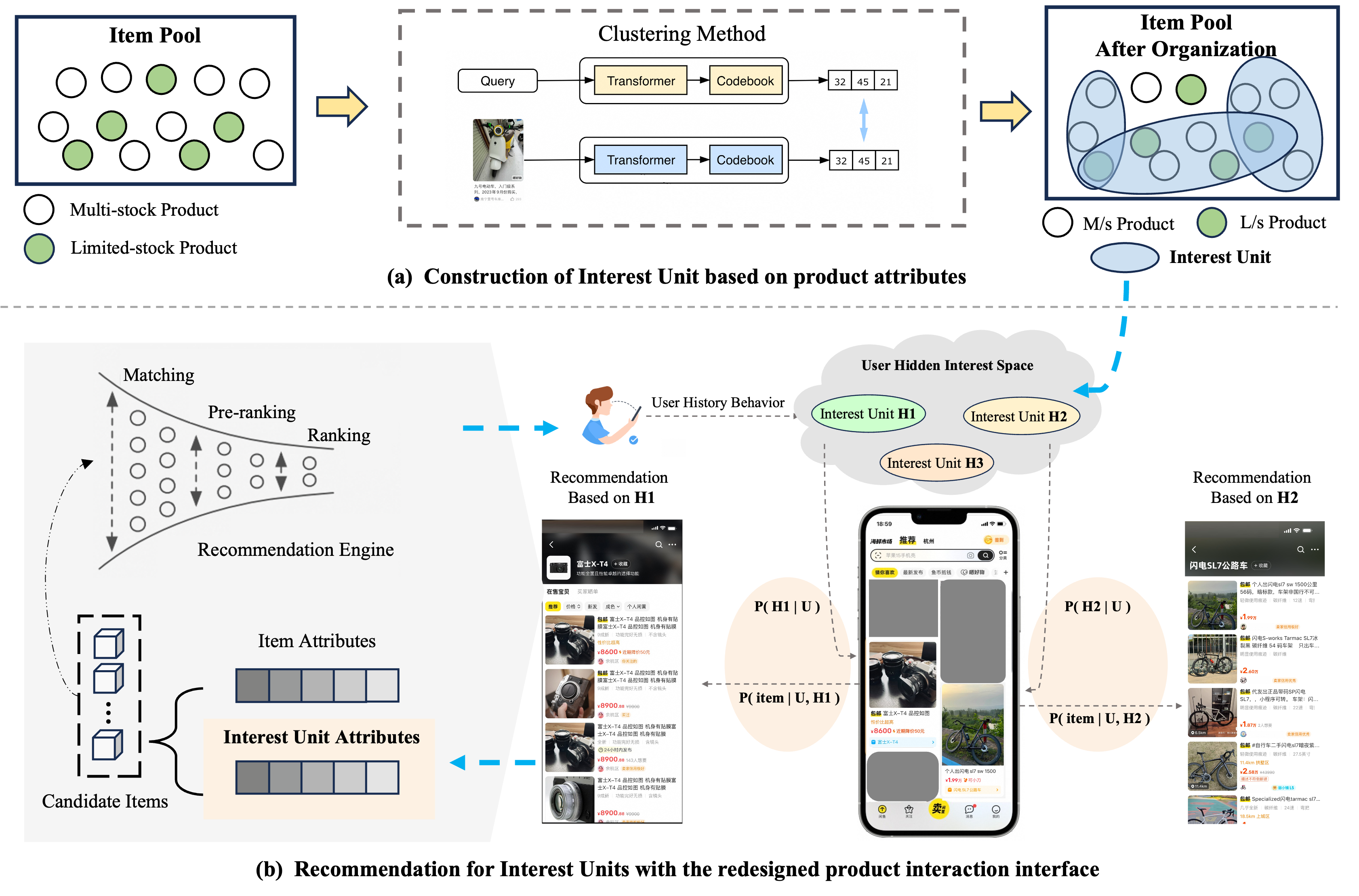}
    \caption{An overview of the IU4Rec framework, which consists of (a) the construction of interest unit: products are organized into Interest Unit using DSI-based clustering methods on top of the product attributes and textual information~\cite{Tay2022TransformerMA}, and (b) the recommendation of interest unit and the corresponding products. More specifically, the latter component operates in two stages -- stage one recommends products utilizing both item and interest unit attributes, and stage two focuses on identifying products in the given interest unit.}
    \label{fig:iu4rec_overview}
\end{figure*}

The personalized recommendation system plays a vital role in e-commerce platforms ~\cite{cvr1, din,youtubednn,tencent_www}, helping users quickly find the products of their interests, and ultimately improving long-term user engagement~\cite{survey_rec,survey_challenges}. Traditional e-commerce platforms predominantly employ historical behaviors to predict user preferences and, meanwhile, apply diversification strategies to mitigate the risk of filter bubbles, ensuring the trade-off between the exploitation of current interests and the exploration of new interests~\cite{Carbonell1998TheUO}.

More specifically, industrial recommendation systems generally adopt the Deep Interest Network (DIN) algorithm for content recommendation~\cite{din}, in which historical user-item interactions were employed to locate the most intriguing items that may interest the user (here, an item refers to a selling product). Although such a historical preference-based recommendation mechanism has been widely explored in industrial systems, it still faces several important challenges, especially for C2C (consumer-to-consumer) platforms like Xianyu. In C2C platforms, products of personal sellers often have limited stock availability~\cite{wu2024metasplit}, and once the item is sold, it is no longer available for distribution. This results in most items distributed on Xianyu having relatively few interactions, affecting the effectiveness of DIN-like recommendation algorithms that largely depend on accumulating user-item interactions. Additionally, those items will eventually be materialized as historical user behaviors, and further influence the recommendation effectiveness.

Common solutions to this challenge include (1) offering traffic support through a separate channel~\cite{wu2022adversarial,poso, cold_cross_domain}, (2) leveraging user behaviors from similar items to enhance the representation of the limited-stock product~\cite{wu2024metasplit,cold_graph,meta_emb}, or (3) employing the debias-based methods to ensure that model training is not dominated by popular items~\cite{logq, class_re}, and thereby, allowing less-exposed items to obtain more generalized representations. 
While these methods can be effective, they do not address the fundamental issue inherent in the item-based recommendation paradigm. In this paradigm, all user interactions are accumulated on top of unique items. Once the item becomes unavailable, the accumulated interactions are wasted.

In addition, we often notice that users tend to browse through a set of similar products on our platform, and then examine each item in detail for the final decision. This behavior stems from the nature of C2C platforms, where most listings by individual sellers include both standard product attributes (like brand and class) and second-hand characteristics (such as usage conditions). Therefore, our recommendation algorithm should capture this process, starting from matching the broad-level interest, and further guiding users to find the best option among similar products within the matched interest. To achieve this goal, we redesign our recommendation process into two stages which leads to the development of IU4Rec, an \textbf{I}nterest \textbf{U}nit-based two-stage \textbf{Rec}ommendation system framework consisting of two main components.

As shown in Figure~\ref{fig:iu4rec_overview}, component (a) outlines the construction of the interest unit. We develop, on top of large language models~\cite{Tay2022TransformerMA}, an effective method to identify interest units and the associated queries and products. By reorganizing products around these interest units, we can persist and accumulate user interactions over time, even when the underlying products are sold out. By design, this approach can effectively aggregate user interests and is appropriate for long-term product distribution on C2C platforms.

Component (b) illustrates the IU-based recommendation process. In addition to upgrading the recommendation algorithm, we extensively redesign the user interface in our system, to make it aligned with the two-stage recommendation process. In stage one, we aim to recommend the interest units rather than individual items. Once the user expresses his/her interest, e.g. by clicking the Interest Unit, the recommendation process turns into the next stage. In stage two, the recommendation system shifts its goal to recommend specific products within the chosen interest unit.

The new interest unit based user interaction design makes it possible for the system to aggregate user interactions at the IU level, and across different users. Consequently, we have enhanced our model with the integration of interest unit, and propose an IU-Boosted Network that leverages both item-level and interest-unit-level features, and utilize hierarchical interest unit click sequences to enhance user interests modeling. 


Compared to traditional recommendation systems that directly estimate product efficiency, we believe the new two-stage recommendation paradigm based on interest units enhances the precision of user interest modeling and improves decision-making efficiency by presenting similar products at a high density. This approach can simplify purchase decisions for C2C users and offer C2C e-commerce platforms a more effective way to gather user interactions at the  same time.


We evaluated our method by conducting experiments on a production dataset and online A/B testing. The results demonstrate the superiority of our proposed IU4Rec paradigm, which is now fully deployed on Alibaba's Xianyu Platform.

To summarize, the main contributions of our work are as follows:
\begin{itemize}
    \item To the best of our knowledge, this is the first solution to optimize recommendation system through item organization, interaction interface design, and recommendation model simultaneously. We propose a novel recommendation paradigm for C2C e-commerce platforms called IU4Rec by grouping similar items into clusters called interest units, redesigning the interaction interface for enhanced platform efficiency and user experience, and propose a novel recommendation model  to enhance CTR prediction ability. We believe this holistic approach opens new possibilities for systematic improvements.
    
    \item The upgraded product format has fundamentally changed user interaction patterns, enabling cross-product behaviors within the same interest unit. We enhance our algorithm through an IU-Boosted Network that aggregates user behaviors under shared interest unit IDs and leverages collective interactions within interest units. This methodology improves recommendation effectiveness within interest domains and normal product domain 
    at the same time.
    
    \item Experiments on industrial datasets and online A/B tests demonstrate the superiority of our proposed IU4Rec paradigm, validating its advantages in both product format design and recommendation modeling. 
\end{itemize}


%% file: tex/2_related_work.tex
\section{Related Work}
This section briefly introduces the progress of another two-stage recommendation paradigm, the cold-start recommendation for new publish product. The cold-start item recommendation problem, can be challenging due to the limited interactions between users and items~\cite{wu2022adversarial}. Cold-start item recommendation is typically divided into two stages by the amount of user interactions. 

In the first stage, the new items have just entered into the system and thus received very limited exposure. At this moment, any existing recommendation models may fail because of the minimal user interactions. To avoid direct competition with existing items, industrial recommendation systems may offer a separate channel only for new items. In this case, high-quality new items only compete with each other, and they can warm up and enter the next stage after obtaining a certain level of interaction. After that, items can obtain simple real-time statistical information or incorporate a limited amount of user and item features for recommendation. The common approaches ~\cite{cold_cv,cold_context,cold_cross_domain,cold_graph,cold_poi} utilize generalization features or explore information across different domains and modalities. The most effective algorithms, of course, are usually the ones that involve introducing additional user-item interaction data from other domains~\cite{zhang2022keep,rec4ad}.

In the second stage, new items may receive a few interactions from the separate channel, And now, they compete directly with the existing items. 
ClassBalance~\cite{class_re} and LogQ~\cite{logq} attempted to enhance the importance of new items during the training process by adjusting the weights of the samples or their corresponding network parameters. 
DropoutNet~\cite{dropoutnet} and POSO~\cite{poso} enhanced the expressive ability and robustness of models for new items by designing specialized networks tailored to their characteristics. There are indeed many other research studies on this topic, we do not list them all here due to the content limit.

%% file: tex/3_preliminaries.tex
\section{Preliminaries}

\begin{figure}[tbp]
    \centerline{\includegraphics[width=8cm]{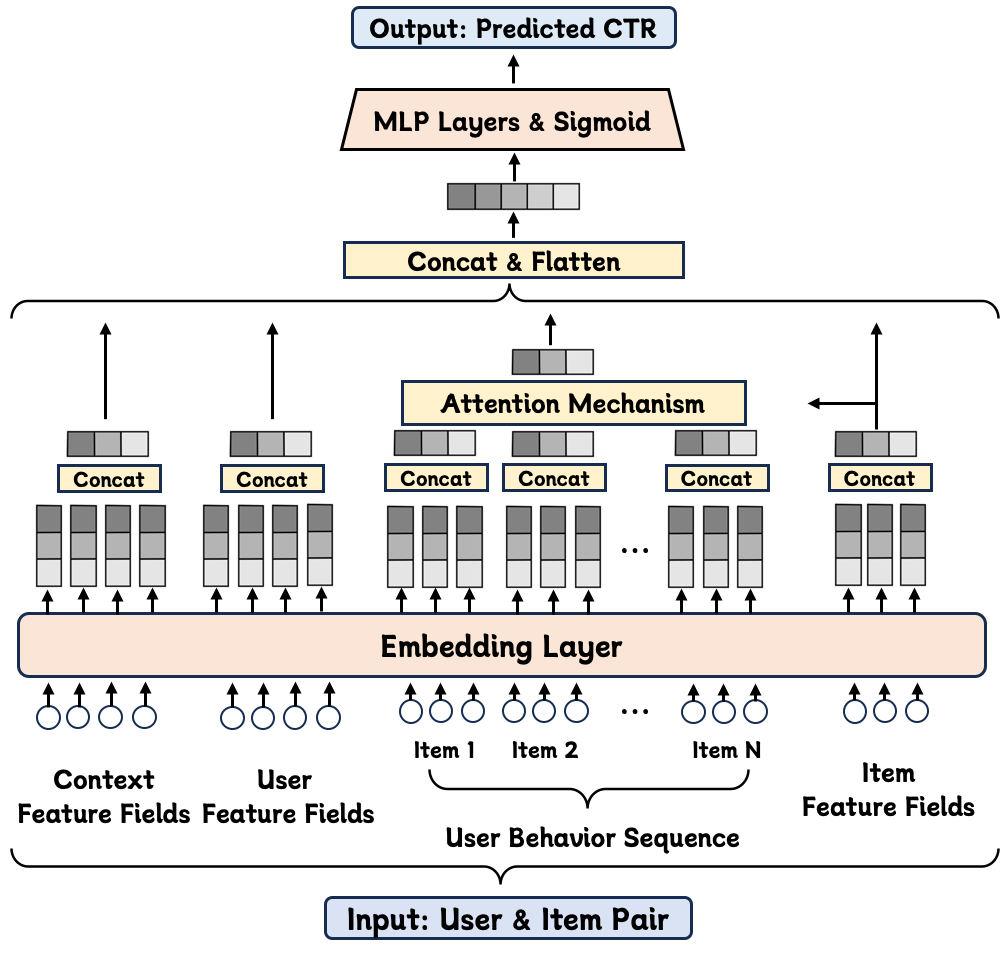}}
    \caption{An illustration of Embedding and MLP structure with sequence modeling for the deep CTR prediction model}
    \label{fig:din_model}
\end{figure}

In this section, we formulate the problem and briefly introduce the existing production CTR prediction model on the large-scale recommender system of Xianyu.

\subsection{Problem Formulation}
Given a dataset $\mathcal{D}=\{ \mathbf{x}, y \}^N,(\mathbf{x}, y)$ marks a sample and $N$ is the number of samples, where $\mathbf{x}$ denotes the high dimensional feature vector consisting of multi-fields (e.g., user and item field), and $y$ is the binary label with $y = 1$ indicating the sample is clicked. Our task is to accurately predict the probability of CTR $p_{ctr} = p(y=1|x)$ for the testing sample $x$.

\subsection{Production CTR Prediction Model}

We select Deep Interest Network \cite{din} as our base model due to its online efficiency and effectiveness, which follows the conventional Embedding and MLP paradigm and utilizes an attention mechanism to model user behavior sequences, as depicted in Figure~\ref{fig:din_model}.

\textit{\textbf{Embedding Layer}}: 
The inputs are composed of non-sequential features (e.g., user ID) and sequential features (e.g., user’s history clicked items). The embedding layer is employed to convert each discrete feature from the raw input into a vector of lower dimensions, by using an embedding look-up table. The embedding of non-sequential features is simply concatenated, whereas for the embedding of sequential features, a sequence information modeling module is used to assemble them into a fixed-size representation.

\textit{\textbf{Sequence Information Modeling}}: 
Deep Interest Network utilizes a local attention mechanism to dynamically capture the user's interests based on the similarity between their historical clicked items and the target item. 
This mechanism allows for the acquisition of a personalized representation of the user's interests since it enables weight pooling of the information in the sequence of varying length.
Additionally, DIN could be further optimized by leveraging the target attention mechanism equations~\cite{BST,transformer,pssa} to replace the original local attention mechanism. In the existing Multi-head Target-attention (MHTA) implementation, the target item $Item_t$ is consider as query $(Q)$ and the history click sequence $\boldsymbol{S}_u$ is considered both as keys $(K)$ and values $(V)$, where $\boldsymbol{S}_u=\left\{Item_1, Item_2, \ldots, Item_H\right\}$ is the set of embedding vectors of items in the user behaviors with length of $H$. 

Specifically, the output of MHTA can be formalized as follows :

\begin{equation}
\text { TargetAttention }(Q, K, V)=\operatorname{softmax}\left(\frac{Q K^{\top}}{\sqrt{d}}\right) V,
\end{equation}
where $Q =W^Q Item_t$, $K=W^K X^{\text {seq }}$, and $V=W^V X^{\text {seq }}$, the linear projections matrices $W^Q \in \mathbb{R}^{d \times d}, W^K \in \mathbb{R}^{d \times d}, W^V \in$ $\mathbb{R}^{d \times d}$ are learn-able parameters and $d$ stands for the dimension of hidden space. The temperature $\sqrt{d}$ is introduced to produce a softer attention distribution for avoiding extremely small gradients. Finally, all non-sequential embedding and transformed user sequential embedding is concatenated with other continuous features together to generate the overall embedding and pass through a deep network to get the final prediction.

\textit{\textbf{Loss}}: 
The objective function used in DIN is the negative log-likelihood function defined as:
\begin{equation}
L=-\frac{1}{N} \sum_{(x, y) \in \mathcal{D}}(y \log f(x)+(1-y) \log (1-f(x))),
\end{equation}
where $D$ is the training set, each sample $x$ is associated with a ground-truth label $y$. The output of the model, denoted as $f(x)$, represents the predicted probability of sample $x$ being clicked.

%% file: tex/4_methodology.tex
\section{methodology}
This chapter introduces the construction of interest units, the redesign of product forms with new interaction interface, and the IU-Boosted CTR prediction model integrating interest unit.

\subsection{Interest Unit-based Product Organization}

\begin{figure}[tbp]
\includegraphics[width=8cm]{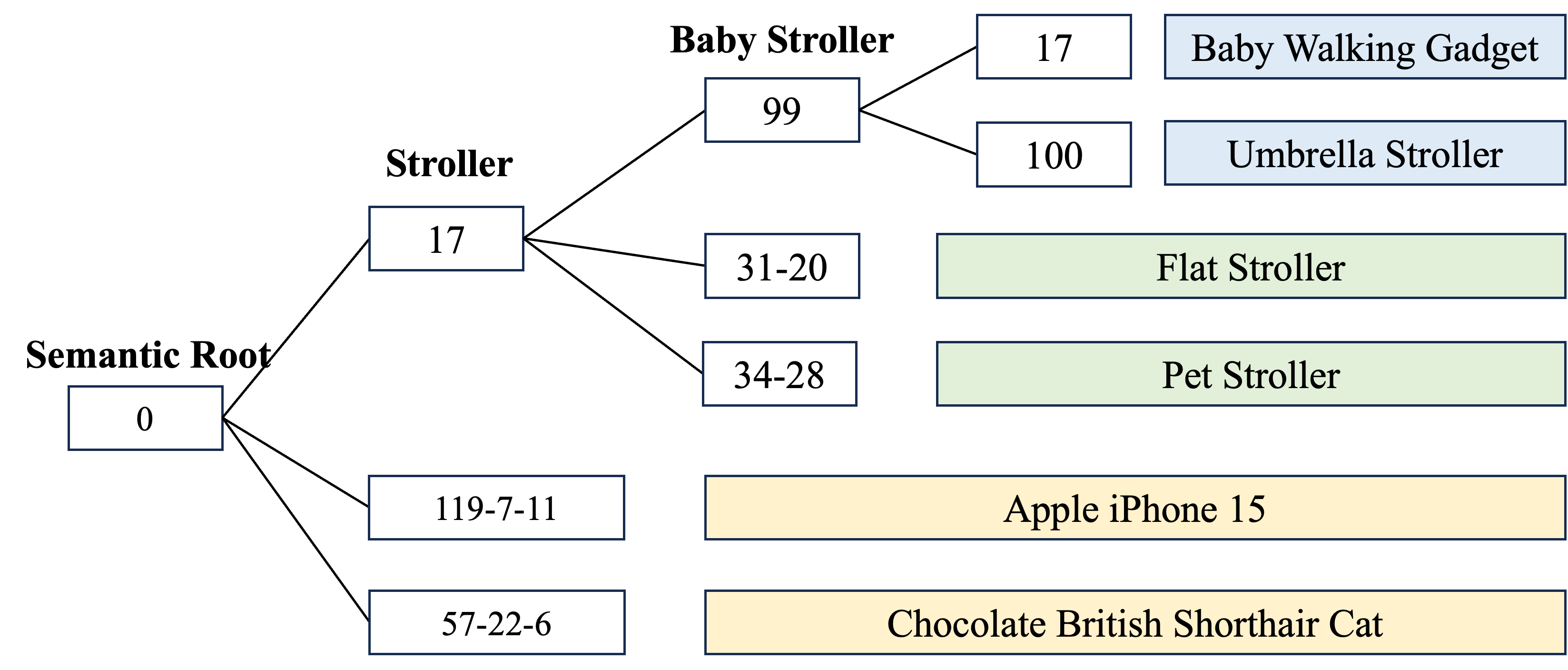}
\caption{One example of the foundational understanding system generated by the semantic clustering method.}
\label{fig:gsid_tree}
\end{figure}

User behaviors such as browsing, searching, clicking, or purchasing are primarily driven by underlying needs, which can be abstracted into specific interest units. These interest units can range from concrete product instances (e.g., "Iphone15 ProMax 256G") to broad demand categories (e.g., "concert tickets"). 
By predefining interest units and systematically associating relevant products with these constructed interest units, platforms can enhance user engaging experience and demand-matching efficiency.

\subsubsection{\textbf{Construction of Interest Unit}} Despite the diverse needs of Xianyu users, we believe that the core demands can be exhaustively identified to some extent. We have adopted a data-driven, bottom-up analytical framework that constructs interest units from the perspectives of product attributes and user needs, reorganizing the vast array of products on the Xianyu platform. 
\\ \textbf{I. Attribute-Driven} \textit{(Based on intrinsic product attributes)} \\
When browsing, users tend to focus on the core attribute information of a product. By combining these core attributes, we can essentially exhaustively identify the core demands users have for a certain category of products. Therefore, based on product attribute information, we have defined two types of user interest units.
\begin{itemize}
    \item SPU Interest Units: Product attributes, such as category and brand, are important on e-commerce platforms. For standard products with comprehensive structured attributes, we aggregate products into SPU Interest units based on the CPV (customer perceived value) information provided by users or identified by algorithms, forming a type of interest unit.
    \item Image Cluster Interest Unit: Product image information is also crucial when users are browsing. For non-standard products where structured attributes are difficult to define, we use product image information to aggregate products with similar appearances into clusters, thereby defining a type of interest unit based on these clusters.
\end{itemize}
\textbf{II. Demand-Aware} \textit{(Balancing product attributes and user needs)}
Not all categories have a one-to-one match between product supply and buyer demand. To balance buyer needs and seller supply during the construction of interest units, we developed a Query-Aware semantic unit generation system called Generative Semantic ID~\footnote{Another systematic effort from industry practice, which isn't the focus of this paper, will be briefly mentioned below. This work will soon be under review.}(GSID), based on open knowledge from large models and combined with the vast interaction data of "query-product" on Xianyu. This system defines Semantic interest units. The Xianyu GSID is a hierarchical tree structure, as shown in the figure~\ref{fig:gsid_tree}. GSID includes three levels, each containing 128 IDs, with the second level space being approximately 16,000 and the third level space being approximately 2.1 million.
Specifically, the Xianyu GSID algorithm uses the encoder-decoder network of the T5 model as the backbone structure. The encoder is a BERT-based vector encoder responsible for extracting product semantic vectors and for the decoder combines the encoder output vector at each decoding step with the previous decoding result to produce the current decoding vector, and then looks up the corresponding theme in the CodeBook to discretize and generate hierarchical semantic IDs.
\subsubsection{\textbf{Redesign of Interaction Interface}}
\begin{figure}[tbp]
\includegraphics[width=8cm]{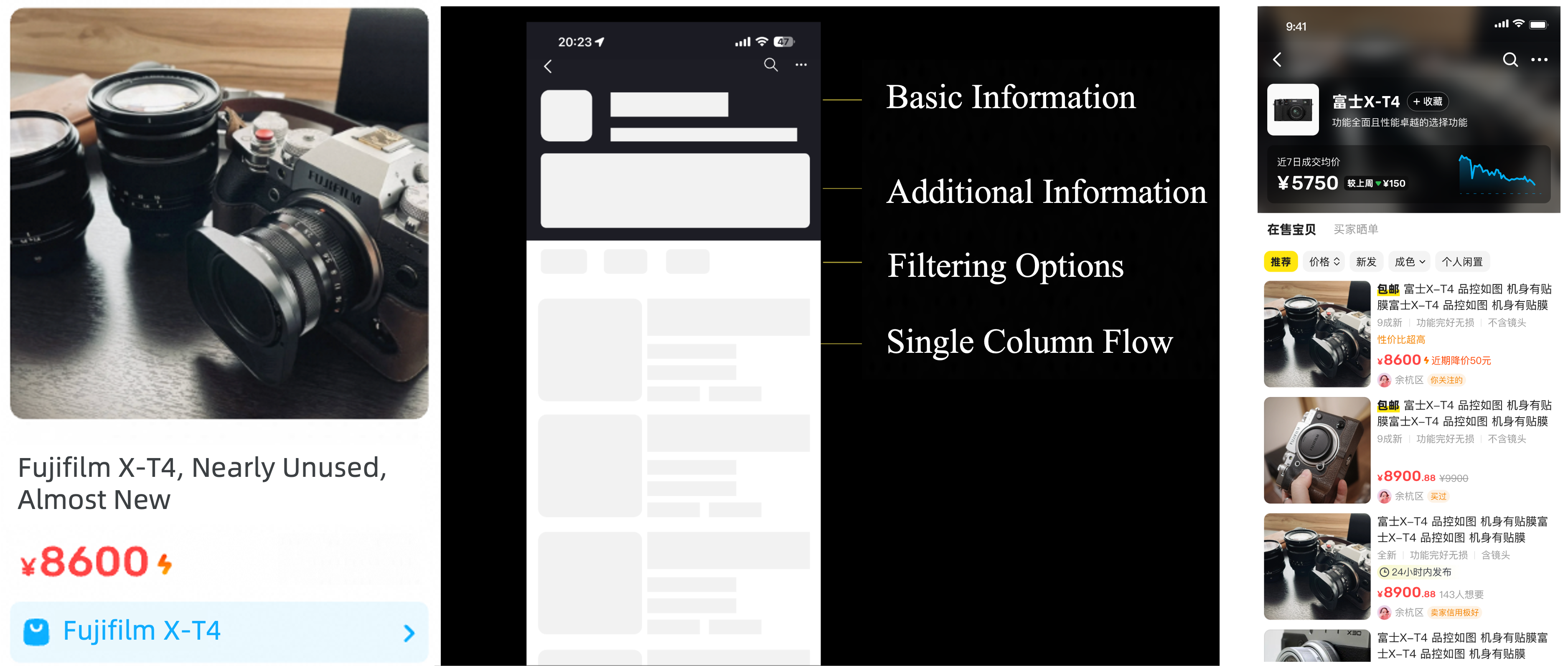}
\caption{The redesigned product format. Left is stage one style  and right image is stage two style with explanationo on the middle}
\label{fig:new_product}
\end{figure}
Once these basic interest units are constructed, they can not only be incorporated into algorithmic modeling but also further utilized to change the way products are presented on the homepage recommendation interface. As shown in Figure \ref{fig:new_product}, we implemented a series of changes to clearly express user interest units: (1) On the homepage recommendations, we display the theme of the associated interest unit next to the product (left side of Figure \ref{fig:new_product}). (2) On the secondary landing page, products within the same interest unit are arranged together, making it easier for users to select products efficiently (right side of Figure \ref{fig:new_product}, product prototype image), with explanations for each module on the secondary landing page in between.
It is noteworthy that this new product format naturally aligns with the aforementioned two-stage recommendation paradigm, allowing for a more organic combination of algorithm models and product formats, significantly enhancing efficiency. Once the product set for interest units is delineated, we need to generate a front-end title for the interest unit to facilitate user understanding. This information is also displayed at the bottom of the card in homepage and at the top of the secondary page. The title of the interest unit is automatically generated by a large language model, by feeding corresponding descriptive information of N products randomly selected from each interest unit.

\subsection{Interest Unit-based Recommendation}\label{sec:recommendation}
\begin{figure*}[tbp]
    \includegraphics[width=14cm]{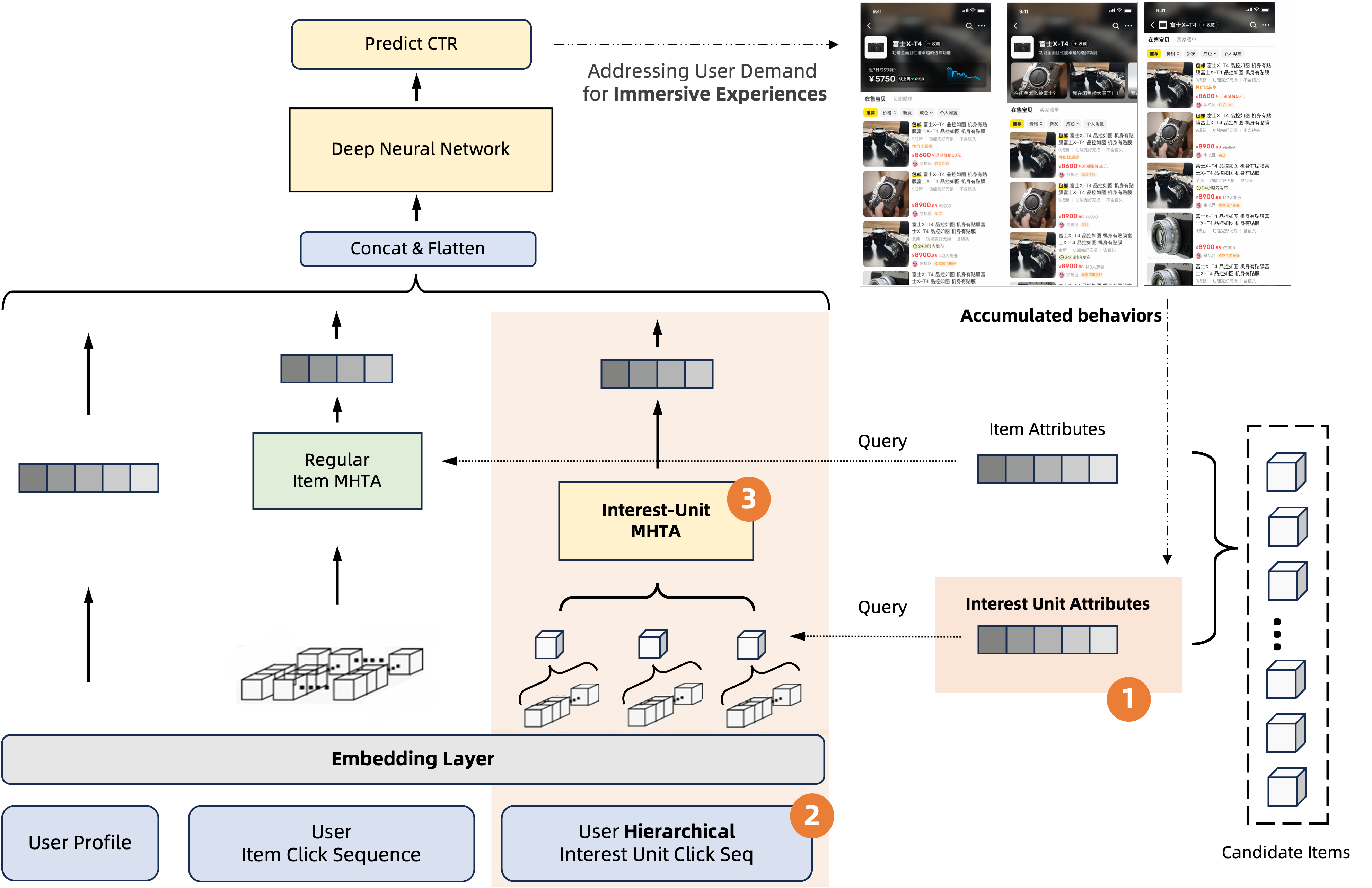}
    \caption{An overview of proposed IU-Boosted Network, which consists of three components: (1) the interest unit-level feature for each product, (2) the user's hierarchical IU click sequence to determine their interest unit preference, and (3) the attention mechanism introduced for handling multiple items within the interest unit.}
    \label{fig:model_overview}
\end{figure*}

As shown in Figure \ref{fig:new_product}, the upgrade in the homepage product format has led to significant changes in user navigation paths: user interactions are no longer confined to individual products but can occur across multiple products under the same interest unit. Additionally, behaviors of different users within the same interest unit can be aggregated and accumulated. 

Building upon this, we construct IU-level features to reflect the attributes of each IU and hierarchical IU click sequences using attention mechanism to user interest unit interest. We name this recommendation algorithm, which leverages behavior accumulation on Interest Unit (IU), as \textbf{IU-Boosted Network}. In this section, we will introduce the components of our proposed method in detail.

\subsubsection{\textbf{IU-Level Feature Construction}}
We accumulate behaviors of different users across all products under the same interest unit to construct IU-Level features, serving as foundational attributes of the interest unit. Products may be deleted after being sold, resulting in the obsolete of the accumulated information on their product IDs. However, the information aggregated on their associated interest unit remains permanently accessible. When new products are launched, we can attach their related interest unit attributes to enhance recommendation efficiency. Based on this, we develop multi-dimensional features to optimize recommendation performance:
(1) Statistical Features IU Dimension: Include various behavioral metrics such as impressions, clicks, inquiries, and transactions, reflecting the overall performance and popularity of the interest unit.
2) User-IU Cross Feature: Capture interaction patterns and frequencies between users and specific interest units or specific types of interest units.
\subsubsection{\textbf{IU Hierarchical Click Sequences}}
Users may exhibit multiple behaviors under the same interest unit, where the number of interactions reflects the intensity of their preference for the interest unit. We construct hierarchical IU click sequences to model user preferences at the interest unit level for refined recommendations. 
The normal item click sequence takes the following form:
\begin{equation}
    \boldsymbol{E}(Item \ Seq)=
    Concat [\boldsymbol E({Item\_i}), i = 1\ldots, m],
\end{equation}
where $\boldsymbol{E}\left({Item}\right)$ means the embedding representation for items consist of ID feature and side feature:
\begin{equation}
    \boldsymbol E\left({Item}\right)=Concat [\boldsymbol E\left(\mathcal{F}_{Item\_ID}\right), \boldsymbol E\left(\mathcal{F}_{Item\_Side}\right)],
\end{equation}

The embedding representation of IU and the IU click sequence can be expressed as followed:
\begin{equation}
    \boldsymbol{E}\left(IU\right)=
    Concat [\boldsymbol E\left(\mathcal{F}_{IU\_ID}\right), \boldsymbol E\left(\mathcal{F}_{IU\_Side}\right),
    \boldsymbol{E}\left(Item \ Seq\right)],
\end{equation}
\begin{equation}
    \boldsymbol{E}\left(IU \ Seq\right)=
    Concat [\boldsymbol E\left({IU\_1}\right), \boldsymbol E\left({IU\_2}\right), \ldots,
    \boldsymbol E\left({IU\_n}\right)],
\end{equation}
where $\boldsymbol{E}\left(IU \ Seq\right)$, $\boldsymbol{E}(IU \ Seq)$ means the sequence embedding representations, and $\boldsymbol E\left(\mathcal{F}_{IU\_ID}\right)$, $\boldsymbol E\left(\mathcal{F}_{IU\_Side}\right)$ means the embedding for id feature and side feature for interest unit respectively.

\subsubsection{\textbf{Attention Mechanism for IU Sequence}}
In addition to the traditional product-based attention mechanism, we further introduce an attention mechanism based on IU behavior. When scoring a target product, we first parse the IU ID associated with the target product and the IU IDs of the products in the user's historical click sequence. We utilize an attention mechanism to calculate the distance between the IU ID of the target product and the IU IDs of products previously clicked on, in IU to assess the intensity of the user's preference for the interest unit to which the current target product belongs.

Please note that our model mainly focuses on the expansion of product attributes from the perspective of single item to interest unit, and thus can be applied to various CTR prediction networks and sequence information modeling methods.

%% file: tex/5_experiments.tex
\section{Experiments}

\begin{table*}[]
\caption{
Performance of constructed interest units, where CTR Improvement is compared with normal product. The real CTR is omitted due to privacy issue.
}
\label{tab:iuanalysis}
\begin{tabular}{ccccccc}
\midrule
\textbf{IU Type\textbackslash Metric} & \textbf{Numbers} & \textbf{Product Coverage} & \textbf{Exposure Ratio} & \textbf{Click Ratio} & 
\textbf{Bills Ratio} & \textbf{CTR Improvement} \\ \toprule
\textbf{SPU}                          & 45.9k            & 23.0\%                    & 9.28\%                  & 10.4\%       & 23.9\%           & +21.92\%                                     \\
\textbf{Image}                        & 287k            & 5.11\%                    & 13.7\%                  & 12.5\%        & 24.3\%        & +34.75\%                                     \\
\textbf{Semantic}                     & 615k            & 78.8\%                    & 77.0\%                  & 77.1\%      & 51.8\%            & +12.85\%                                     \\ \midrule
\end{tabular}
\end{table*}

\begin{table*}[]
\caption{
Model comparison on the production dataset, where "RI" is short for "RelaImpr". The improvements are statistically significant (i.e. two-side t-test with $p < 0.05$) over the original model.
}
\label{tab:offline_result}
\begin{tabular}{cc|cccc|cc|cc}
\midrule
\multicolumn{2}{c|}{\multirow{2}{*}{}}                                                                                                             & \multicolumn{4}{c|}{\textbf{Overall}}                 & \multicolumn{2}{c|}{\textbf{Intereste Unit}} & \multicolumn{2}{c}{\textbf{Normal Product }} \\ \cline{3-10} 
\multicolumn{2}{c|}{  \diagbox{\textbf{Method}}{\textbf{Metric}}  }                                                                                                                      & \textbf{AUC}    & \textbf{RI}   & \textbf{GAUC}   & \textbf{RI}   & \textbf{AUC}                & \textbf{RI}                & \textbf{AUC}                   & \textbf{RI}                       \\ \midrule
\multicolumn{1}{c|}{\multirow{4}{*}{\begin{tabular}[c]{@{}c@{}}Common \\ DNNs\end{tabular}}} & DNN                                                 & 0.7348          & -0.75\%        & 0.6509          & -0.46\%        & 0.6972                      & -1.12\%                   & 0.7428                         & -0.53\%               \\
\multicolumn{1}{c|}{}                                                                        & Wide\&Deep\cite{wideanddeep} & 0.7346          & -0.83\%       & 0.6506          & -0.66\%     & 0.6970                       & -1.22\%                   & 0.7425                         & -0.66\%               \\
\multicolumn{1}{c|}{}                                                                        & DeepFM\cite{deepfm}          & 0.7335          & -1.30\%         & 0.6504          & -0.79\%         & 0.6965                      & -1.47\%                   & 0.7415                         & -1.07\%               \\
\multicolumn{1}{c|}{}                                                                        & DIN\cite{din}                & 0.7366          & 0.00\%        & 0.6516          & 0.00\%        & 0.6994                      & 0.00\%                    & 0.7441                         & 0.00\%                \\ \midrule
\multicolumn{1}{c|}{\multirow{3}{*}{Gate Methods}}                                            & GateNet\cite{gatenet}              & 0.7356          & -0.41\%        & 0.6512          & -0.26\%         & 0.6984                      & -0.52\%                    & 0.7432                         & -0.37\%               \\
\multicolumn{1}{c|}{}                                                                        & FiBiNet\cite{senet}                & 0.7355          & -0.45\%         & 0.651           & -0.40\%         & 0.6982                      & -0.62\%                    & 0.7431                         & -0.41\%               \\
\multicolumn{1}{c|}{}                                                                        & POSO\cite{poso}                    & 0.7365          & -0.03\%         & 0.6515          & -0.07\%         & 0.6988                      & -0.32\%                    & 0.744                          & -0.04\%               \\ \midrule
\multicolumn{1}{c|}{\multirow{2}{*}{Graph Methods}}                                           & GroupEmb\cite{groupid}             & 0.7367          & 0.05\%          & 0.6517          & 0.07\%          & 0.6995                      & 0.04\%                    & 0.7443                         & 0.08\%                \\
\multicolumn{1}{c|}{}                                                                        & GIFT\cite{gift}                    & 0.7368          & 0.10\%          & 0.652           & 0.26\%          & 0.6999                      & 0.24\%                    & 0.7445                         & 0.16\%                \\ \midrule
\multicolumn{1}{c|}{\begin{tabular}[c]{@{}c@{}}Former \\ Baseline\end{tabular}}              & MSNet\cite{wu2024metasplit}        & 0.7392          & 1.11\%          & 0.6544          & 1.85\%         & 0.7039                      & 2.25\%                    & 0.7468                         & 1.11\%                \\ \midrule
\multicolumn{1}{c|}{\textbf{Ours}}                                                           & IU-Boosted Network                                               & \textbf{0.7411} & \textbf{1.91\%} & \textbf{0.6549} & \textbf{2.17\%} & \textbf{0.7082}             & \textbf{4.39\%}            & \textbf{0.7484}                & \textbf{1.76\%}                 \\ \midrule
\end{tabular}
\end{table*}

\begin{table}[]
\caption{Comparison between production model and IU-Boosted Model in two domains during online A/B tests followed by industry standards. 
}
\begin{tabular}{cccc}
\toprule
  \diagbox{\textbf{Domain}}{\textbf{Metric}}     & \textbf{CTR} & \textbf{Clicks} & \textbf{Bills}  
                     \\ \toprule
\textbf{Overall} & +2.90\%      & +2.62\%         & +1.02\%                     \\
\textbf{Interest Unit Rec} & +11.76\%      & +10.22\%         & +5.61\%                    \\
\textbf{General Product Rec} & +1.58\%      & +1.49\%         & +0.33\%        \\ \toprule
\end{tabular}
\label{tab:online result}
\end{table}

In this section, we conduct extensive experiments on the offline
dataset and online A/B testing to evaluate the performance of the proposed method. The following research questions (RQs) are addressed:
\begin{itemize}
    \item \textbf{RQ1}: How do the constructed interest units perform, and have they achieved the desired outcomes in terms of efficiency and coverage?  (Section 5.2)
    \item \textbf{RQ2}: How does the new product interaction interface perform and how do the metics compared with former product interface? (Section 5.2)
    \item \textbf{RQ3}: Does your proposed IU-Boosted Network outperform state-of-the-art performance? (Section 5.3)
    \item \textbf{RQ4}: Does your proposed model work in real large-scale online recommendation scenarios? (Section 5.2 and 5.4)
\end{itemize}
\subsection{Experiment Setup}

\subsubsection{\textbf{Xianyu Dataset}}
We collect click traffic logs of 8 days from Alibaba’s Xianyu Recommendation Platform\footnote{The proposed paradigm and the proposed method are implemented from a industrial practice and there are no suitable public data-set.} to build the production data-set with 10 billion samples (1.5 billion sample per day), with numbers of items and people, 209 features (e.g., user and item features). 
The samples in the first 7 days and 8th day are employed for training and testing respectively, and we randomly partition the testing set into 10 parts and report average evaluation results. 

\subsubsection{\textbf{Baselines}}
We compare the proposed IU-Boosted Network with three types of baselines: the first type are common DNN-based methods, including:
\begin{itemize}
    \item \textbf{DNN}: a method proposed by Youtube, and has been been implemented in various industrial applications.
    \item \textbf{Wide \& Deep} \cite{wideanddeep}:  a method proposed by Google that consists of a wide model and a deep model and has gained significant adoption in various applications.
    \item \textbf{DeepFM} \cite{deepfm}: a method proposed by Google that combines a wide model with a deep model and has been widely adopted in various applications.
    \item \textbf{DIN} \cite{din}: our base model, as described in Section 4, which uses attention mechanism to  aggregate historical behaviors sequence information based on the similarity between historical behaviors item and the target item.
\end{itemize}
The second type are some gate network methods.
\begin{itemize}
    \item \textbf{GateNet} \cite{gatenet}: a method proposed by Baidu, using gate layer to generate weights for the embedding layer and the hidden layer.
    \item \textbf{FiBiNet} \cite{senet}: a method proposed by Sinanet, filtering the information of embedding based on their importance using the SENet module.
    \item \textbf{POSO} \cite{poso}: a method proposed by KuaiShou, a short-video application. We implement it as described in the paper with personal MLP and personal MHA.
\end{itemize}
The third type are graph learning methods.
\begin{itemize}
    \item \textbf{GroupID}~\cite{groupid}: a meta-learning approach proposed by Airbnb to generate new item embeddings by calculating the average of similar item embeddings. 
    \item \textbf{GIFT}~\cite{meta_emb}: a meta-learning approach to address the cold-start problem by generating ID embeddings for cold items using other available features.
\end{itemize}

\subsubsection{\textbf{Evaluation Metric}}
Following previous works \cite{din,dien,star,clid}, we adopt some widely used metrics, AUC, GAUC and RelaImpr, to evaluate the performance of our proposed method. 
The calculation of AUC, GAUC, RelaImpr can be expressed as follow:
\begin{equation}
\mathrm{AUC}=\frac{1}{|P||N|} \Sigma_{p \in P} \Sigma_{n \in N} I(\Theta(p)>\Theta(n)),
\end{equation}
where $P$ and $N$ denote positive sample set and negative sample set, respectively. $\Theta$ is the estimator function and $I$ is the indicator function.

\begin{equation}
\mathrm{GAUC}=\frac{\sum_{i=1}^n \# \text { impression }_i \times \mathrm{AUC}_i}{\sum_{i=1}^n \text { \#impression }_i},
\end{equation}
where the AUC is first calculated within samples of each user, and averaged w.r.t sample count, where $n$ is the number of users, \#impression $i$ and $\mathrm{AUC}_i$ are the number of impressions and AUC corresponding to the $i$-th user.

\begin{equation}
\text { RelaImpr }=\left(\frac{\text { AUC }(\text { measured model })-0.5}{\text { AUC }(\text { base model })-0.5}-1\right) \times 100 \%
\end{equation}
where RelaImpr metric is used to measure relative improvement over models and 0.5 stands for the AUC of a random guesser. Generally, higher values for AUC and GAUC indicate better performance.

\subsubsection{\textbf{Implementation Details.}}
We implement the DNN part of DeepFM, Wide \& Deep, DNN, DIN all the same architecture, i.e., a three-layer MLP Network with 512, 256 and 128 hidden units. For all attention layers in above models, the number of hidden units are all set to 128. AdagradDecay optimizer is adopted in all the methods, the learning
rate of 1e-4 is set. Batch size is set to 4096 for all DNNs. And the historical click sequence is collected within the last 30 days and max length is 150. The max length of the click sequence inside IU click sequence is 5.

\subsection{RQ1 \& RQ2: Product Format Comparison and Analysis of Constructed Interest Unit}
After the launch of the two-stage recommendation and distribution based on interest units, we chose to integrate it into the homepage in the form of a workflow. This means the homepage has two workflows, one for distributing normal product recommendations and the other for IU product recommendations (workflow in Figure~\ref{fig:iu4rec_overview} and product format in Figure~\ref{fig:new_product}). Both are merged into a single queue based on certain rules~\footnote{These rules are mainly established from a business decision perspective} and presented to users. To verify the effectiveness of this product format, we conducted an A/B test with 10\% of the traffic. The experimental group used the aforementioned product format, while the control group used the original product recommendation format. The experiment lasted for 7 days, and the new product format achieved a 5\% improvement in bills, a 10\% improvement in GMV, and even in exposure, there was a minor decrease in clicks. Using 13\% of the exposure, it contributed to 18\% of the GMV, proving the effectiveness of this product format. We rapidly expanded this product format to a larger traffic volume, leaving a reversal bucket for observation.
The exposure ratio of IU on the homepage exceeded 10\%. The constructed types of interest units are presented in Table~\ref{tab:iuanalysis}, and we can see the click-through rate for different IUs have certain advantages compared to the overall product efficiency.

\subsection{RQ3: Model Comparison With Baselines}
To demonstrate the effectiveness and superiority of the proposed method, we conduct offline experiments on the Xianyu dataset and compare it with state-of- the-art (SOTA) approaches. The experimental results are presented in Table~\ref{tab:offline_result}. 
Compared with previous approaches, our method achieved the best overall performance. Specifically, when compared to the baseline model DIN, our approach delivered a 1.91\% enhancement in AUC and a 2.17\% increase in GAUC~\footnote{Note that the 0.1\% AUC gain is already considerable in large-scale industrial recommendation system \cite{chang2023pepnet}}
. Additionally, our method achieved the highest AUC metrics for both interest units and normal products recommendation.

\subsection{RQ4: Online A/B test}
In Xianyu's recommendation system, we compared a new model's effectiveness against a highly-optimized baseline (DIN architecture) through an online A/B test.
The 7-day experiment period provided statistically validated, reliable results. We used CTR, Clicks, and Bills  as metrics to evaluate the online effectiveness.
The results are shown in Table~\ref{tab:online result}.
The IU-Boosted model demonstrated significant improvements in both interest unit and normal product recommendations. Specifically, for interest unit recommendations, there was an 11.76\% increase in CTR, a 10.22\% rise in Clicks, and a 5.61\% boost in Bills. For normal product recommendations, CTR improved by 3.72\% and Clicks by 5.20\%. Overall, the model achieved a 2.90\% increase in CTR, a 2.62\% rise in Clicks, and a 1.02\% increase in Bills, highlighting a notable performance enhancement.

%% file: tex/6_conclusions.tex
\section{Conclusion}
In this study, we introduce a new paradigm of interest unit-based recommendation algorithms, IU4Rec, aimed at optimizing recommendation performance on C2C platforms and better capturing user-product interactions. 
We construct interest units based on product attributes and user needs, and redefine the product interaction interface to let the interaction remain on the interest unit and product at the same time. Furthermore, we implement the IU-Boosted Network to improve CTR prediction, IU4Rec effectively addresses the challenge of utilizing behavior data from sold-out products and tackles the limited-stock issue inherent to our C2C platform.
The proposed IU4Rec has been proven effective through experiments and online A/B testing and is now fully deployed on the Xianyu Platform, enhancing the experience for millions of users daily.

Although this study primarily focuses on the C2C platform, the above practice can be easily extended to other platforms. Consequently, we hope that the proposed IU4Rec paradigm can offer valuable insights to researchers and practitioners in the field of recommendation systems.